\UseRawInputEncoding
\documentclass[letterpaper]{IEEEtran}

\ifCLASSINFOpdf

\else

\fi

\usepackage{flushend}
\usepackage{stfloats}
\usepackage{subfigure}
\usepackage{fancyhdr}
\usepackage{mathrsfs}
\usepackage{graphicx}
\usepackage{geometry}
\usepackage{color}
\usepackage{placeins}
\usepackage{float}
\usepackage{tabularx,colortbl}
\usepackage{amsmath}
\usepackage{amsfonts}
\usepackage{amssymb}
\usepackage{amsthm}
\usepackage{bm}
\usepackage{cases}
\usepackage{subeqnarray}
\usepackage[numbers,sort&compress]{natbib}
\usepackage[ruled,linesnumbered]{algorithm2e}
\begin{document}

\title{Worst-case Design for RIS-aided Over-the-air Computation with Imperfect CSI}
\author{\IEEEauthorblockN{Wenhui~Zhang, Jindan~Xu\IEEEauthorrefmark{0}, \IEEEmembership{Member,~IEEE}, Wei~Xu\IEEEauthorrefmark{0}, \IEEEmembership{Senior Member,~IEEE}, \\Xiaohu~You\IEEEauthorrefmark{0}, \IEEEmembership{Fellow,~IEEE}, and Weijie~Fu\IEEEauthorrefmark{0}}
\vspace{-1cm}
\thanks{W. Zhang, J. Xu, W. Xu, and X. You are with the National Mobile Communications Research Laboratory (NCRL), Southeast University, Nanjing 210096, China (\{whzhang, jdxu, wxu, xhyou\}@seu.edu.cn).

W. Fu is with Guangdong Communications and Networks Institute, Guangdong 518131, China (fujw@gdcni.cn).
}
}

\maketitle
\thispagestyle{fancy}
\renewcommand{\headrulewidth}{0pt}
\pagestyle{fancy}
\cfoot{}
\rhead{\thepage}

\newtheorem{mylemma}{Lemma}
\newtheorem{mytheorem}{Theorem}
\newtheorem{mypro}{Proposition}
\begin{abstract}
Over-the-air computation (AirComp) enables fast wireless data aggregation at the receiver through concurrent transmission by sensors in the application of Internet-of-Things (IoT). To further improve the performance of AirComp  under unfavorable propagation channel conditions, we consider the problem of computation distortion minimization in a reconfigurable intelligent surface (RIS)-aided AirComp system. In particular, we take into account an additive bounded uncertainty of the channel state information (CSI) and the total power constraint, and jointly optimize the transceiver (Tx-Rx) and the RIS phase design from the perspective of worst-case robustness by minimizing the mean squared error (MSE) of the computation. To solve this intractable nonconvex problem, we develop an efficient alternating algorithm where both solutions to the robust sub-problem and to the joint design of Tx-Rx and RIS are obtained in closed forms. Simulation results demonstrate the effectiveness of the proposed method.
\end{abstract}

% keywords

% \fi
\IEEEpeerreviewmaketitle
\vspace{-0.3cm}
\section{Introduction}

%\IEEEPARstart{T}{he} development of Internet-of-Things (IoT) is in full swing, involved in all aspects of our daily lives.
\IEEEPARstart{A}{s} the most prominent characteristic of the developing information age, massive data with astounding growth is of critical significance in facilitating the Internet-of-Things (IoT) applications. However, it has become harder for data collection from increasing smart devices of IoT. Meanwhile, it is even more challenging for information fusion of massive data, also known as data aggregation, from the numerous sensor devices with limited spectrum and low latency constraint.

To address this issue, a promising solution, namely over-the-air computation (AirComp), was proposed by leveraging the superposition property of wireless multiple access channels (MAC) \cite{Airinfo}. Through concurrent transmissions by sensors and a weighted average function for the distributed local computing, fast data aggregation can be achieved by AirComp at the receiver, realizing a low transmission latency that is independent of the number of IoT devices. Recently, AirComp has attracted much attention in many areas like information theory \cite{Airinfo}, signal processing \cite{Airsig2}, and transceiver design \cite{Airtran}. It has been proved that AirComp can effectively improve communication efficiency and reduce the required bandwidth \cite{Airperf}. However, the advantages of the AirComp are promised upon that the propagation channel conditions are favorable.

To solve this problem, reconfigurable intelligent surface (RIS), emerging as a complementary technology \cite{QWuTowards}, is believed to improve signal propagation conditions by reflecting incident signals at favorable angles utilizing the passive reflecting elements \cite{RISoptim}\cite{RIS222}. A RIS assisted AirComp system is proposed in \cite{RISAir0} to boost the received signal power and mitigate the performance bottleneck by reconfiguring the propagation channels. Thus with the assistance of using the RIS, energy focusing and nulling at desired locations can be realized in a AirComp system, shaping favorable propagation links for data collection. To reduce the mean squared error (MSE) of AirComp, the authors in \cite{IRSAir} developed an alternating difference-of-convex (DC) programming algorithm by optimizing the transceiver and the RIS phase shifts. Besides, the authors in \cite{IRSAir2} tackled the non-convex optimization problem with the matrix lifting and concave convex procedure, solving the DC problem and designing the RISs¡¯ phases and a linear detector for the RIS-aided cloud radio access network (C-RAN).

However, the assumption of perfect CSI in these studies for AirComp is hard to achieve in practice, resulting in noticeable performance degradation due to CSI inaccuracy in practical scenarios, not to mention the worst-case condition. To the best of our knowledge, one exception is \cite{CSIerror2}, which proposed an efficient iterative algorithm to obtain a suboptimal solution for maximizing the system sum-rate considering the robustness against the impact of CSI imperfection. In addition, the wireless sensors of an AirComp system are usually powered by limited-power beacons \cite{AirPower}, leading to a sum-power constraint of all the sensors.

Compared with these existing studies, in this paper we consider the optimization problem of transceiver design and RIS phase selection in a RIS-aided AirComp system under the sum-power constraint with imperfect CSI. Aiming to minimize the computation MSE under the total power constraint with CSI uncertainty, we cope with the joint design problem under the worst-case robustness for the AirComp. To solve this complicated nonconvex problem, we fortunately obtain closed-form solution to the subproblem of the robust design, and derive the closed-form solutions to the joint design of transceivers and RIS phases in an alternating optimization manner.

\emph{Notations}: Throughout this paper, ${(\cdot)}^{\rm H}$, ${(\cdot)}^{\rm T}$, ${(\cdot)}^*$, and ${(\cdot)}^{\dagger}$ denote the conjugate transpose, transpose, conjugate, and Moore-Penrose inverse, respectively. $\mathbb{R}(\mathbb{C})$ and $\mathbb{E}\{\cdot\}$ represent the space of real (complex) numbers and the expectation operator. $\left|\cdot\right|$ denotes the modulus of a scalar. $\mathcal{CN}(0,1)$ represents the distribution of a circularly symmetric complex Gaussian variable with zero mean and unit variance. $U[0,2\pi)$ represents the uniform distribution drawn from 0 to 2$\pi$. The $k$th entry of vector $\boldsymbol x$ and the $(i,j)$th element of matrix $\mathbf{X}$ are represented by $[\boldsymbol x]_{k}$ and $[\mathbf{X}]_{ij}$, respectively.
\vspace{-0.5cm}
\section{System Model and Problem Formulation}

\begin{figure}
\centering
\includegraphics[width=3.45in]{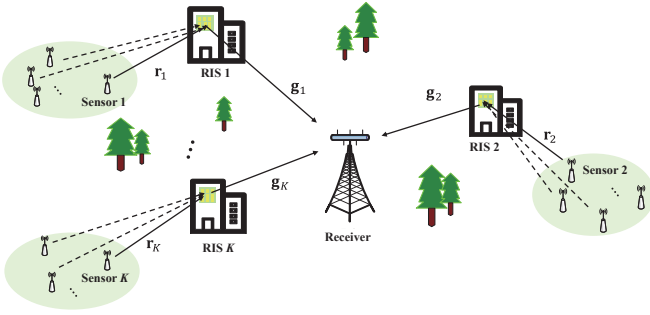}
\vspace{-0.5cm}
\caption{Over-the-air computation with RIS.}
\vspace{-0.5cm}
\label{Fig:block diagram}
\end{figure}

We consider the uplink of a RIS-assisted AirComp system, which is composed of $K$ single-antenna sensors and one single-antenna receiver. As shown in Fig. 1, assume that the direct links of sensor-receiver are blocked due to unfavorable propagation conditions, such as large obstacles \cite{QWuTowards}\cite{RISCE}, and are thus neglected. Each RIS with $N$ passive reflecting antenna elements helps the communication from the sensors in a specific serving area to the receiver. Since the geometric locations of these RISs are relatively far from each other, the signal from one sensor is dominantly reflected by the RIS in the serving area, while propagations via the other RISs are ignorably weak due to large path loss. In a certain time slot, the RIS in each area only provides auxiliary reflection communication service for data uploading from a single scheduled sensor.

%Each sensor pre-processes its own measurement signal and then sends it to the receiver through a MAC channel simultaneously via the RIS reflections. Through the AirComp, waveform superposition property of a wireless channel can be utilized to realize the aggregation of data simultaneously transmitted by devices \cite{Airperf}.
\vspace{-0.4cm}
\subsection{Processing of Air Computation}

In the RIS-assisted AirComp system, each sensor pre-processes its own measurement signal and then sends it to the receiver for further calculation. Let ${ x}_{k} \in \mathbb{R}$, $\forall k \in \{1,...,K\}$ denote the pre-processed original signal of sensor $k$, which without loss of generality is assumed to have a normalized variance. Let $\psi_{k}:\mathbb{R}\rightarrow \mathbb{R}$, $\forall k \in \{1,...,K\}$ denote a pre-processing function of sensor $k$ and function $\varphi:\mathbb{R}\rightarrow \mathbb{R}$ denotes the receiver's post-processing operation. Then, the received computation signal at the receiver is expressed as
\begin{equation}\label{eq:yo}
y=\varphi \bigg(\sum_{k=1}^{K}\psi_{k}(x_{k})\bigg).
\end{equation}

From \eqref{eq:yo}, by the pre-processing operation $\psi_{k}$ by each sensor and the post-processing operation $\varphi$ at the receiver, the receiver obtains the desired signal summation directly instead of requesting one-by-one transmissions of  $x_{k}$ and a specific computation \cite{Airequa}. By integrating computation and communication exploiting the signal superposition property of the wireless multiple-access channel, AirComp is accomplished via this concurrent transmission.

As to practical applications, for example, a central controller collects and computes the average real-time data (e.g., velocity, acceleration, temperature, humidity) sent by distributed sensors. By post-processing the operator like an averaging function of $\varphi$, the overall state of the environment is acquired.
\vspace{-0.5cm}
\subsection{Computation  Model}

In this system, each sensor data is linearly scaled by a factor  $t_{k} \in \mathbb{C}$ before transmission, realizing the pre-processing operation  $\psi_{k}$. And the received signal is linearly scaled by a Rx-scaling factor  $m$ realizing the post-processing operation $\varphi$. Applying this to \eqref{eq:yo}, the received signal becomes
\begin{align}\label{eq:y}
y&=m\bigg(\sum_{k=1}^{K}\mathbf{g}_{k}^{\rm H}\mathbf{\Phi}_{k}\mathbf{r}_{k}t_{k}x_{k}+n\bigg)\nonumber\\
&=m\bigg(\sum_{k=1}^{K}\underbrace{\mathbf{g}_{k}^{\rm H}\mathrm{diag}\{\mathbf{r}_{k}\}}_{\mathbf{h}_{k}^{\rm H}}\boldsymbol{v}_{k}t_{k}x_{k}+n\bigg),
\end{align}
where $\mathbf{g}_{k}\in\mathbb{C}^{N \times 1}$ and $\mathbf{r}_{k}\in\mathbb{C}^{N \times 1}$ denote the channel between the $k$th RIS and the receiver and the channel between the $k$th sensor and its corresponding RIS, respectively. The diagonal matrix $\mathbf{\Phi}_{k}=\mathrm{diag}(\beta_{k,1}\mathrm{e}^{j\theta_{k,1}},\cdots,\beta_{k,N}\mathrm{e}^{j\theta_{k,N}}) \in \mathbb{C}^{N \times N}$ is defined as the reflection matrix of RIS $k$ and $\boldsymbol{v}_{k}\triangleq[\beta_{k,1}\mathrm{e}^{j\theta_{k,1}},\cdots,\beta_{k,N}\mathrm{e}^{j\theta_{k,N}}]^{\mathrm{T}} \in\mathbb{C}^{N \times 1}$ is a vector reshaped from the diagonal elements of $\mathbf{\Phi}_{k}$, where $\beta_{k,i} \in [0,1]$, is the amplitude coefficient of the $i$th reflecting element of RIS $k$ and $\theta_{k,i}\in (0,2\pi]$ is the phase coefficient for $ i\in\{1,2,\cdots, N\}$. In this paper, we assume $\beta_{k,i} = 1$ without loss of generality. Scalar $n \in \mathbb{C}$ is the additive white Gaussian noise (AWGN) following $\mathcal{CN}(0,\sigma^2)$, and $\mathbf{h}_{k}^{\rm H} \triangleq \mathbf{g}_{k}^{\rm H}\mathrm{diag}\{\mathbf{r}_{k}\} \in \mathbb{C}^{1 \times N}$ is the cascaded equivalent channel. Especially in \eqref{eq:y}, the sum power of all the $K$ sensors in the AirComp system is limited to ${P} \triangleq \sum_{k=1}^{K}{\vert t_{k}\vert}^2$ for energy saving.

Then, the computation distortion in terms of the AirComp MSE, as defined in \cite{AirPower}, is evaluated as
\begin{align}
{\rm MSE}&\triangleq\mathbb{E}\bigg\{\left| y-\sum_{k=1}^{K}x_{k}\right|^{2}\bigg\}\nonumber\\
&=\sum_{k=1}^{K}{\left| m\mathbf{h}_{k}^{\rm H}\boldsymbol{v}_{k}t_{k}-1\right|}^{2}+\sigma^2{\vert m\vert}^{2}.
\end{align}

Note that the MSE depends on the instantaneous CSI which can be obtained by channel estimation \cite{RISCE,RISCSI1} rather than the stochastic nature of $\mathbf{g}_{k}$ and $\mathbf{r}_{k}$. However, the obtained CSI is always noisy, which follows the deterministic CSI error model \cite{CSIerror} and is mathematically formulated as
\begin{equation}\label{eq:delta}
\Delta_{\mathbf{h}_{k}}=\mathbf{h}_{k}^{\rm H}-\hat{\mathbf{h}}_{k}^{\rm H},
\end{equation}
where $\mathbf{h}_{k}$, for $\forall k\in\{1,2,\cdots, K\}$, is the nominal channel for the $K$ links, and the CSI error $\Delta_{\mathbf{h}_{k}}$ is bounded , i.e.,
\begin{equation}\label{eq:CSIbound}
\|\Delta_{\mathbf{h}_{k}}\|_{2} \leq \varepsilon,
\end{equation}
which means that the CSI uncertainty region is confined within a region of radius $\varepsilon$. By substituting \eqref{eq:delta}, the MSE in \eqref{eq:y} conditioned on the CSI estimate,  $\hat{\mathbf{h}}_{k}$,  is rewritten as
\begin{align}\label{eq:mse}
{\rm MSE}=\sum_{k=1}^{K}{\left| m(\hat{\mathbf{h}}_{k}^{\rm H}+\Delta_{\mathbf{h}_{k}})\boldsymbol{v}_{k}t_{k}-1\right|}^{2}+\sigma^2{\vert m\vert}^{2}.
\end{align}

\vspace{-0.5cm}
\subsection{Problem Formulation}

The aim of our work is to minimize the AirComp MSE in \eqref{eq:mse} subject to the sum power constraint under the worst-case CSI error by jointly optimizing the design of the transceiver and the RISs' reflection matrices. Now we are ready to formulate the optimization problem as follows
\begin{subequations}\label{eq:p1}
\begin{align}
\min_{m,\{t_k\},\{\boldsymbol{v}_{k}\}} \max_{\Delta_{\mathbf{h}_{k}^{\rm H}}}\quad &\sum_{k=1}^{K}\left| m(\hat{\mathbf{h}}_{k}^{\rm H}+\Delta_{\mathbf{h}_{k}})\boldsymbol{v}_{k}t_{k}-1\right|^{2}+\sigma^2\vert m\vert^2\nonumber\\
\\
{\rm s.t.}\quad
&\sum_{k=1}^{K}\vert t_k \vert^2 \leq P,\label{eq:pconstraint}\\
&\vert \boldsymbol{v}_{k}(n)\vert^{2}=1,\quad \forall n\in\{1,\cdots,N\},\label{eq:psub22}\\
&\|\Delta_{\mathbf{h}_{k}}\|_{2} \leq \varepsilon.\label{eq:deltacons}
\end{align}
\end{subequations}
In particular, \eqref{eq:pconstraint} indicates the sum power constraint of all the $K$ sensors, \eqref{eq:psub22} describes the unit-modulo constraints of the phase-shift element of RIS, and \eqref{eq:deltacons} corresponds to the CSI error bound in (5). Observing this optimization problem, we find that it is hard to obtain the globally optimal solution, especially when coping with the nonconvex objective under the nonconvex unit-modulus constraint.

%In particular, \eqref{eq:pconstraint} indicates the sum power constraint of all the $K$ sensors in the considered AirComp system, \eqref{eq:psub22} describes the unit-modulo constraints of the phase-shift element of RIS, and \eqref{eq:deltacons} corresponds to the CSI error bound in (5) due to CSI uncertainty. Observing this optimization problem, we find that it is hard to obtain the globally optimal solution, especially when coping with the nonconvex objective under the nonconvex unit-modulus constraint.}}
\vspace{-0.3cm}
\section{Proposed Worst-case Design for RIS-aided AirComp System}

In this section, we tackle the optimization problem in \eqref{eq:p1} by jointly optimizing the design of the scaling factors $m,\{t_k\}$ and the RISs' reflection matrices $\{\boldsymbol{v}_{k}\}$ with the goal of minimizing the worst-case MSE in \eqref{eq:p1}. We adopt the idea of alternative optimization, which is to alternately fix some variables while optimizing the others. We start with the robust design with $m,\{t_k\}$ and $\{\boldsymbol{v}_{k}\}$ fixed, which allows us to facilitate the problem solving and fortunately gives a closed-form solution for this worst-case solution.
\vspace{-0.4cm}
\subsection{Robust Design for the Worst-Case CSI Error}

Assuming that $m,\{t_k\}$ and $\{\boldsymbol{v}_{k}\}$ are fixed, problem \eqref{eq:p1} is equal to
\begin{subequations}
\begin{align}\label{eq:pa1}
\max_{\Delta_{\mathbf{h}_{k}}, \forall k}\quad &\sum_{k=1}^{K}\left| m(\hat{\mathbf{h}}_{k}^{\rm H}+\Delta_{\mathbf{h}_{k}})\boldsymbol{v}_{k}t_{k}-1\right|^{2}+\sigma^2\vert m\vert^2\\
{\rm s.t.}\quad&\|\Delta_{\mathbf{h}_{k}}\|_{2} \leq \varepsilon.
\end{align}
\end{subequations}
Then, the above problem in \eqref{eq:pa1} can be decoupled into $K$ subproblems as follows. For each $k \in \{1,2,\cdots,K\}$, we have
\begin{subequations}\label{eq:p22}
\begin{align}
\min_{\Delta_{\mathbf{h}_{k}}}\quad &-\left| mt_{k}(\hat{\mathbf{h}}_{k}^{\rm H}+\Delta_{\mathbf{h}_{k}})\boldsymbol{v}_{k}-1\right|^{2}\\
{\rm s.t.}\quad&\|\Delta_{\mathbf{h}_{k}}\|_{2}-\varepsilon \leq 0.
\end{align}
\end{subequations}

\begin{mytheorem}
The optimal solutions of $\Delta_{\mathbf{h}_{k}}$, and the optimized objective of problem \eqref{eq:p22} are given by the closed-form expressions as
\begin{align}
%&\lambda_{k} ={\rm max}\big\{\vert\hat{t}_{k}\vert^2N- \sqrt{\frac{\vert\hat{t}_{k}\vert^2N\vert \hat{t}_{k}\hat{\mathbf{h}}_{k}^{\rm H}\boldsymbol{v}_{k}-1\vert^2}{\varepsilon^2}},\nonumber\\
%&\quad\quad\quad \vert\hat{t}_{k}\vert^2N+\sqrt{\frac{\vert\hat{t}_{k}\vert^2N\vert \hat{t}_{k}\hat{\mathbf{h}}_{k}^{\rm H}\boldsymbol{v}_{k}-1\vert^2}{\varepsilon^2}}\big\}\label{eq:lambda}\\
%&\lambda_{k} =\vert\hat{t}_{k}\vert^2N- \sqrt{\frac{\vert\hat{t}_{k}\vert^2N\vert \hat{t}_{k}\hat{\mathbf{h}}_{k}^{\rm H}\boldsymbol{v}_{k}-1\vert^2}{\varepsilon^2}},\label{eq:lambda}\\
&\Delta_{\mathbf{h}_{k}}=\frac{\vert\hat{t}_{k}\vert^2 \hat{\mathbf{h}}_{k}^{\rm H}\boldsymbol{v}_{k}\boldsymbol{v}_{k}^{\rm H}-\hat{t}_{k}\boldsymbol{v}_{k}^{\rm H}}{\lambda_{k}-\vert\hat{t}_{k}\vert^2N},\label{eq:deltahk}\\
&{\rm MSE}^{\star}=\sum_{k=1}^{K}\left|\frac{\hat{t}_{k}\hat{\mathbf{h}}_{k}^{\rm H}\boldsymbol{v}_{k}-1}{1-\lambda^{-1}_{k} \vert\hat{t}_{k}\vert^2\boldsymbol{v}_{k}^{\rm H}\boldsymbol{v}_{k}}\right|^2+\sigma^2{\vert m\vert}^{2},\label{eq:msestar}
\end{align}
where $\hat{t}_{k} \triangleq mt_{k}$.
\end{mytheorem}
\begin{IEEEproof}
See Appendix A.
\end{IEEEproof}

This theorem successfully copes with the worst-case CSI uncertainty in the system design.
\vspace{-0.4cm}
\subsection{Joint RIS Reflection and Transceiver Design}

From the above subsection, we obtain the optimal solution in a closed form which achieves the worst-case MSE. Observing \eqref{eq:msestar} and using $\boldsymbol{v}_{k}^{\rm H}\boldsymbol{v}_{k}=N$, it is equivalent to rewrite the problem in \eqref{eq:msestar} as
\begin{align}\label{eq:pa6}
\min_{\{\hat{t}_{k}\},\{\boldsymbol{v}_{k}\}}\quad&\sum_{k=1}^{K}\bigg|\frac{\hat{t}_{k}\hat{\mathbf{h}}_{k}^{\rm H}\boldsymbol{v}_{k}-1}{1-\lambda^{-1}_{k}\vert\hat{t}_{k}\vert^2N}\bigg|^2+\sigma^2{\vert m\vert}^{2}\nonumber\\
{\rm s.t.}\quad&\sum_{k=1}^{K}\vert t_k \vert^2 \leq P,\nonumber\\
&\vert \boldsymbol{v}_{k}(n)\vert^{2}=1, \forall n\in\{1,\cdots,N\}.
\end{align}
It is obvious that the sum power constraint is active since the larger $t_k$ is, the smaller the MSE is. Then, it is safe to rewrite the constraint in \eqref{eq:pa6} as the equality constraint, i.e., $\sum_{k=1}^{K}\vert t_k \vert^2 = P$. The problem in \eqref{eq:pa6} becomes
\begin{align}\label{eq:pa7}
\min_{\{\hat{t}_{k}\},\{\boldsymbol{v}_{k}\}}\quad&\sum_{k=1}^{K}\bigg|\frac{\hat{t}_{k}\hat{\mathbf{h}}_{k}^{\rm H}\boldsymbol{v}_{k}-1}{1-\lambda^{-1}_{k}\vert\hat{t}_{k}\vert^{2}N}\bigg|^2+\frac{\sigma^2}{P}\sum_{k=1}^{K}{\vert\hat{t}_{k}\vert}^{2}\nonumber\\
{\rm s.t.}\quad&\vert \boldsymbol{v}_{k}(n)\vert^{2}=1, n\in\{1,\cdots,N\}.
\end{align}

Note that the problem in \eqref{eq:pa7} is equivalent to the original problem in \eqref{eq:p1}, but with respect to a reduced number of optimization variables including $t_k$ and $m$. However, the problem is still less tractable due to the unit-modulus nonconvex constraints and the coupling variables. Since the problem in \eqref{eq:pa7} can be decoupled into $K$ subproblems for the independent variables with subscript $k$, without loss of generality, we focus on the $k$th subproblem and simplfy the MSE objective in \eqref{eq:msestar} as
\begin{equation}\label{eq:MSE11}
{\rm MSE}_{k} \triangleq \bigg|\frac{\hat{t}_{k}\hat{\mathbf{h}}_{k}^{\rm H}\boldsymbol{v}_{k}-1}{1-\lambda^{-1}_{k}\vert \hat{t}_{k} \vert^{2}N}\bigg|^2+\frac{\sigma^2}{P}{\vert \hat{t}_{k}\vert}^{2}.
\end{equation}

Then we apply the idea of alternating optimization and temporarily relax the unit-modulus constraint. Firstly, for any fixed $m$ and $t_k$, we minimize the MSE by solving the following equality
\begin{equation}
\frac{\partial ({\rm MSE}_{k})}{\partial \boldsymbol{v}_{k}^{*}}=\frac{\vert \hat{t}_{k} \vert^2\hat{\mathbf{h}}_{k}\hat{\mathbf{h}}_{k}^{\rm H}\boldsymbol{v}_{k}-\hat{t}_{k}^{*}\hat{\mathbf{h}}_{k}}{\big({1-\lambda^{-1}_{k}N\vert \hat{t}_{k} \vert^2}\big)^{2}}=0,
\end{equation}
which yields
\begin{equation}\label{eq:vk}
\boldsymbol{v}_{k}=\frac{1}{\hat{t}_{k}}\big(\hat{\mathbf{h}}_{k}\hat{\mathbf{h}}_{k}^{\rm H}\big)^{\dagger}\hat{\mathbf{h}}_{k}.
\end{equation}
Considering the unit-modulus constraint which is temporarily relaxed, we can extract the phase parameters of \eqref{eq:vk} by normalizing the amplitude, and thus the unit-modulus solution for $\boldsymbol{v}_{k}$ becomes irrelevant to $\hat{t}_{k}$. Then for the solution of $\hat{t}_{k}$, we have the following theorem.

\begin{mytheorem}
The optimal solutions for the scaling factors $m$ and $t_k$, equivalently $\hat{t}_{k}$ in \eqref{eq:pa6} are given as
\begin{align}
&m=\sqrt{\frac{1}{P}\sum_{k=1}^{K}\vert \hat{t}_{k} \vert^2},\label{eq:m2}\\
&t_k=\hat{t}_{k}\sqrt{\frac{P}{\sum_{k=1}^{K}\vert \hat{t}_{k} \vert^2}},\label{eq:tk2}
\end{align}
where $\vert \hat{t}_{k} \vert^2=\frac{\sqrt[3]{\frac{2\lambda^{-1}_{k}NP\vert\hat{\mathbf{h}}_{k}^{\rm H}\big(\hat{\mathbf{h}}_{k}\hat{\mathbf{h}}_{k}^{\rm H}\big)^{\dagger}\hat{\mathbf{h}}_{k}-1\vert^2}{\sigma^2}}+1}{\lambda^{-1}_{k}N}$.
\end{mytheorem}
\begin{IEEEproof}
By substituting \eqref{eq:vk}, the \eqref{eq:MSE11} becomes
\begin{equation}\label{eq:MSE22}
{\rm MSE}_{k} \triangleq \bigg|\frac{\hat{\mathbf{h}}_{k}^{\rm H}\big(\hat{\mathbf{h}}_{k}\hat{\mathbf{h}}_{k}^{\rm H}\big)^{-1}\hat{\mathbf{h}}_{k}-1}{1-\lambda^{-1}_{k}\vert \hat{t}_{k} \vert^{2}N}\bigg|^2+\frac{\sigma^2}{P}{\vert \hat{t}_{k}\vert}^{2}.
\end{equation}

Observing the ${\rm MSE}_k$in \eqref{eq:MSE22}, it is easy to check that
\begin{equation}
\frac{\partial^{2}{\rm MSE}_{k}}{\partial (\vert \hat{t}_{k} \vert^2)^2}=\frac{6(\lambda^{-1}_{k}N)^2\vert\hat{\mathbf{h}}_{k}^{\rm H}\big(\hat{\mathbf{h}}_{k}\hat{\mathbf{h}}_{k}^{\rm H}\big)^{-1}\hat{\mathbf{h}}_{k}-1\vert^2}{(1-\lambda^{-1}_{k}N\vert \hat{t}_{k} \vert^2)^4} > 0,
\end{equation}
which allows us to obtain the optimal solution by forcing the following derivative to zero. It follows
\begin{equation}
\frac{\partial{\rm MSE}_{k}}{\partial (\vert \hat{t}_{k} \vert^2)}=\frac{2\lambda^{-1}_{k}N\vert\hat{\mathbf{h}}_{k}^{\rm H}\big(\hat{\mathbf{h}}_{k}\hat{\mathbf{h}}_{k}^{\rm H}\big)^{\dagger}\hat{\mathbf{h}}_{k}-1\vert^2}{(1-\lambda^{-1}_{k}N\vert \hat{t}_{k} \vert^2)^3}+\frac{\sigma^2}{P}=0,
\end{equation}
and thus
\begin{equation}
\vert \hat{t}_{k} \vert^2=\frac{\sqrt[3]{\frac{2\lambda^{-1}_{k}NP[\hat{\mathbf{h}}_{k}^{\rm H}\big(\hat{\mathbf{h}}_{k}\hat{\mathbf{h}}_{k}^{\rm H}\big)^{\dagger}\hat{\mathbf{h}}_{k}-1]^2}{\sigma^2}}+1}{\lambda^{-1}_{k}N}.
\end{equation}
Using the equalities $mt_k=\hat{t}_{k}$ and $\sum_{k=1}^{K}\vert t_k\vert^2=P$, we are able to get the closed-form solutions in \eqref{eq:m2} and \eqref{eq:tk2} for $m$ and $t_k$, respectively.
\end{IEEEproof}

Now we have obtained the optimal closed-form solutions for $m$ in \eqref{eq:m2} and $\{t_k\}$ in \eqref{eq:tk2}. By using \eqref{eq:vk}, we also  obtain the solution for the RISs' reflection matrices $\{\boldsymbol{v}_{k}\}$.  In summary, the proposed solution for solving the original problem in \eqref{eq:p1} is described in Algorithm 1.

Note that the computational complexity of the proposed algorithm for solving problem (7) is dominated by the computations of matrix inversions and matrix multiplications. Concretely, the complexity of the calculations in $\|\Delta_{\mathbf{h}_{k}}\|^2=\varepsilon^2$ is $\mathcal{O}(N)$, and the complexity of the calculations in (16) is $\mathcal{O}(N^3)$. For calculating (17) and (18), the dominating complexity is the calculation of the intermediate variables $\hat{t}_{k}$, which is $\mathcal{O}(N^3)$. Then for each iteration in Algorithm 1, the overall computational complexity amounts to $\mathcal{O}(KN^3)$.

\begin{algorithm}[t]
\LinesNumbered
\SetKwRepeat{Do}{do}{while}
\caption{Proposed algorithm for solving \eqref{eq:p1}}\label{algorithm}
\KwIn{Threshold $\delta > 0$}
initialization: $n = 1$\;
\For{$k$ \KwTo $K$}{
initialize $\boldsymbol{v}_{k}^{(1)}$ satisfying \eqref{eq:psub22}\;
$m^{(1)}=1$, $t_k^{(1)}=P/K$\;
}
\Do{$\sum_{k=1}^{K}\|\boldsymbol{v}_{k}^{(n)}-\boldsymbol{v}_{k}^{(n-1)}\|_2^2+\vert t_k^{(n)}- t_k^{(n-1)} \vert^2+\vert m^{(n)}-m^{(n-1)} \vert^2 \leq \delta$}
{
\For{$k$ \KwTo $K$}{
Calculate $\lambda_{k}^{(n)}$ by solving the equation $\|\Delta_{\mathbf{h}_{k}}^{(n)}\|^2=\varepsilon^2$\;
Update $\boldsymbol{v}_{k}^{(n+1)}$ according to \eqref{eq:vk} and normalize the amplitude\;
Update $m^{(n+1)}$ according to \eqref{eq:m2}\;
Update $t_k^{(n+1)}$ according to \eqref{eq:tk2}\;
}
$n=n+1$\;
}
\KwOut{$m=m^{(n-1)}$, $\{t_k\}=\{t_k^{(n-1)}\}$, $\{\boldsymbol{v}_{k}\}=\{\boldsymbol{v}_{k}^{(n-1)}\}$}

\end{algorithm}

\vspace{-0.3cm}
\section{Simulation Results}

In this section, we present the numerical results for the proposed algorithm.  We define the CSI error level, $\varepsilon=s\left\|\mathbf{h}_{k}\right\|_{2}$, where $s$ represents the CSI uncertainty coefficient which is a metric for evaluating the relative amount of the CSI uncertainty. For simulation, $s$ is set as 0.4 and 0.6, and the channel coefficients $\sigma^2_{h}$ is set as 0.5. We evaluate the performance in terms of the normalized mean squared error (NMSE) of the computation distortion at the receiver and the signal-to-noise ratio (SNR) is defined as $10{\rm log}_{10}(P/\sigma^2_{\rm n})$.

Fig. 2 evaluates the effectiveness of our proposed scheme compared with the non-robust scheme in \cite{AirPower}, where $N=16$, $K=10$ and $P=10$. It is obvious that the proposed method achieves better performance than the non-robust scheme. Meanwhile, CSI uncertainty coefficient $s$ is correlated with the computation MSE, indicating that CSI error confined even within a smaller region results in lower MSE. In addition, we exploit the method by using many random initial values to acquire the globally optimum (approximate) in a numerical way with tractable complexity. It is encouraging that the performance of the proposed method is close to the performance of a globally optimal solution.
%, which validates the necessity of robust design when CSI uncertainty exists

Fig. 3 compares the computation NMSE at the receiver with different numbers of  RIS reflecting elements,  where $K=8$, $s=0.4$, and $P=100$. It is observed that with increasing RIS reflecting elements, the NMSE gradually improves. The reason is  that with the growth of the RIS reflecting elements which can be regarded as more antennas, an extra channel gain is provided. However, with the increase of the RIS elements, the cost and power consumption problem will not be ignored, making it critical for the tradeoff design.

Fig. 4 evaluates the effect of the number of sensor devices on the performance of the system with $N=64$, $P=100$, and $s=0.4$. As we can see in Fig. 4, the performance deteriorates when more and more sensors are getting accessed. With an increasing number of sensors under the total power constraint, the performance gain of the proposed robust design becomes more pronounced.

%\begin{figure}[t]
%\centering
%\includegraphics[width = 0.45\textwidth]{PaperFig/0523final_s04s06opts04.eps}
%\vspace{-0.4cm}
%\caption{The NMSE versus the CSI error level.}
%\vspace{-0.4cm}
%\label{Fig:NMSEepsilon}
%\end{figure}
%
%\begin{figure}[t]
%\centering
%\includegraphics[width =0.45\textwidth]{PaperFig/semilogA0215_N16_512_K8_P100_SNR_0_9_h5_m1_s04.eps}
%\vspace{-0.4cm}
%\caption{The NMSE versus the number of RIS reflecting elements.}
%\vspace{-0.4cm}
%\label{Fig:NMSEantenna}
%\end{figure}
%
%\begin{figure}[t]
%\centering
%\includegraphics[width = 0.45\textwidth]{PaperFig/semilogA0216_N64_K4_14_P100_SNR_0_9_h5_m1_s04.eps}
%\vspace{-0.4cm}
%\caption{The NMSE versus the number of sensor devices.}
%\vspace{-0.4cm}
%\label{Fig:NMSEp}
%\end{figure}

\begin{figure*}[t]
\centering
\begin{minipage}{0.3\linewidth}
\includegraphics[width=2.1in,height=1.73in]{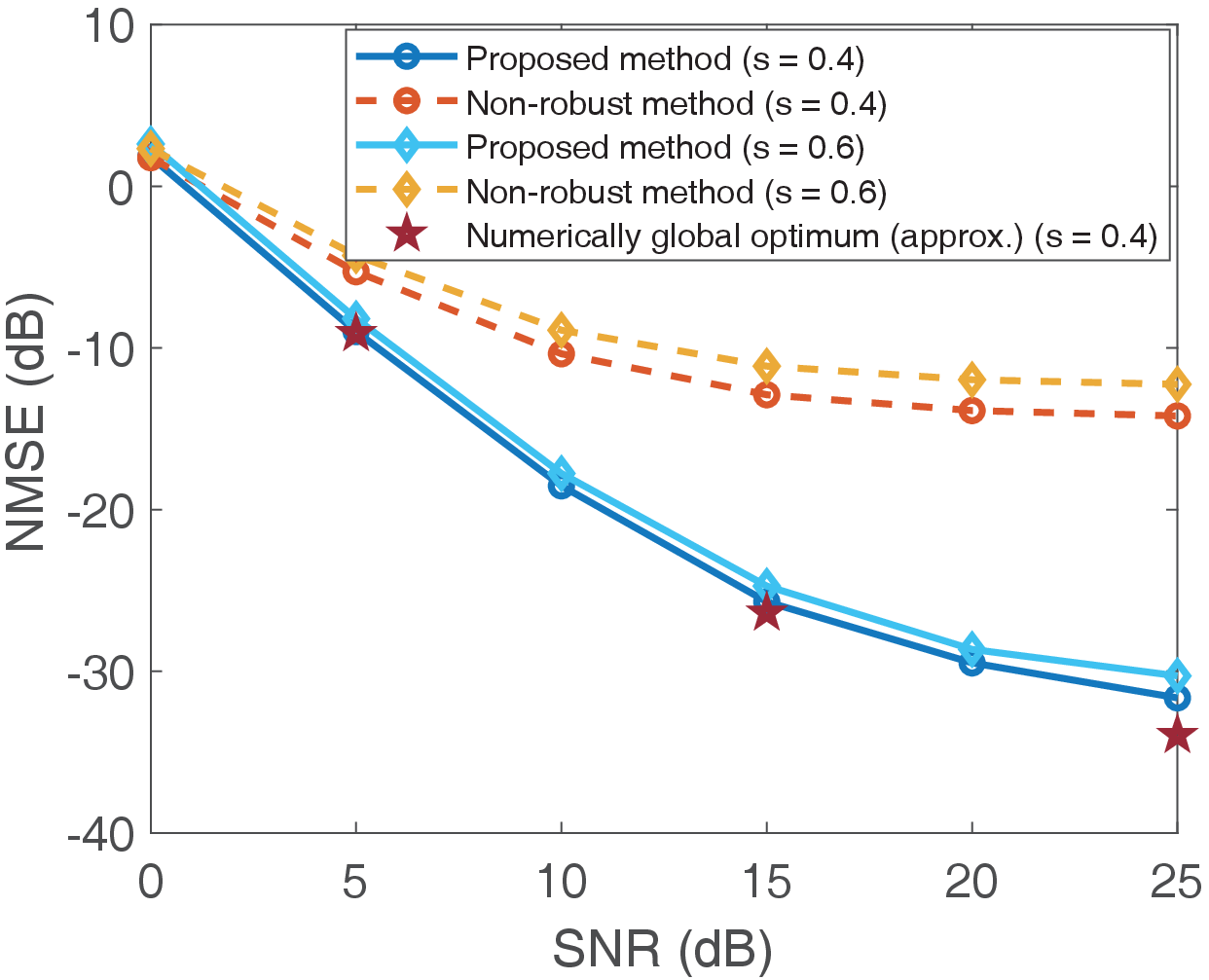}
\vspace{-0cm}
%\caption{The NMSE versus the CSI error level.}
\caption{The NMSE versus SNR.}
\vspace{-0.4cm}
\label{Fig:NMSEepsilon}
\end{minipage}%
\begin{minipage}{0.3\linewidth}
\centering
\includegraphics[width=2.15in,height=1.73in]{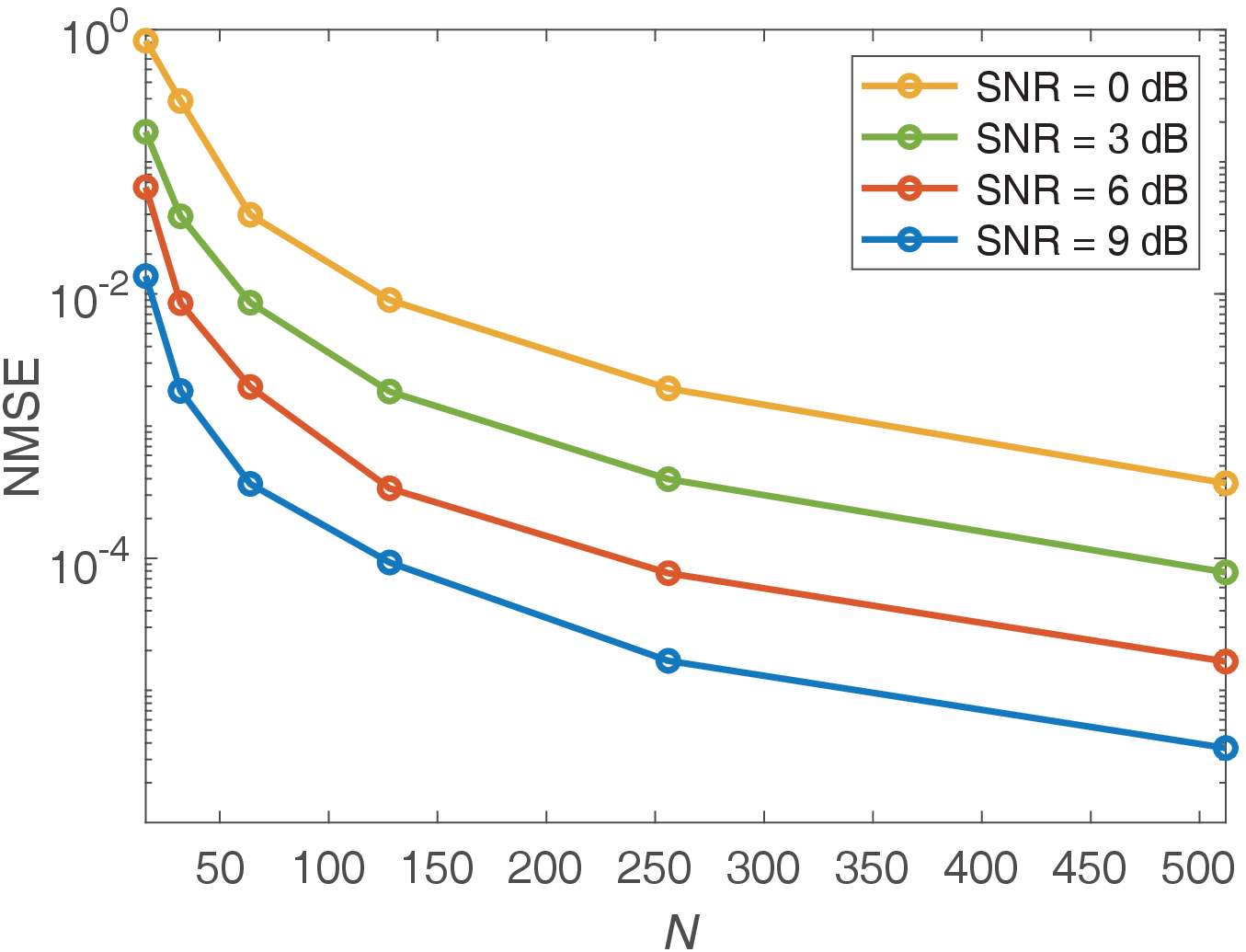}
\vspace{-0.5cm}
%\caption{The NMSE versus the number of RIS reflecting elements.}
\caption{The NMSE versus $N$.}
\vspace{-0.4cm}
\label{Fig:NMSEantenna}
\end{minipage}
\begin{minipage}{0.3\linewidth}
\centering
\includegraphics[width=2.15in,height=1.7in]{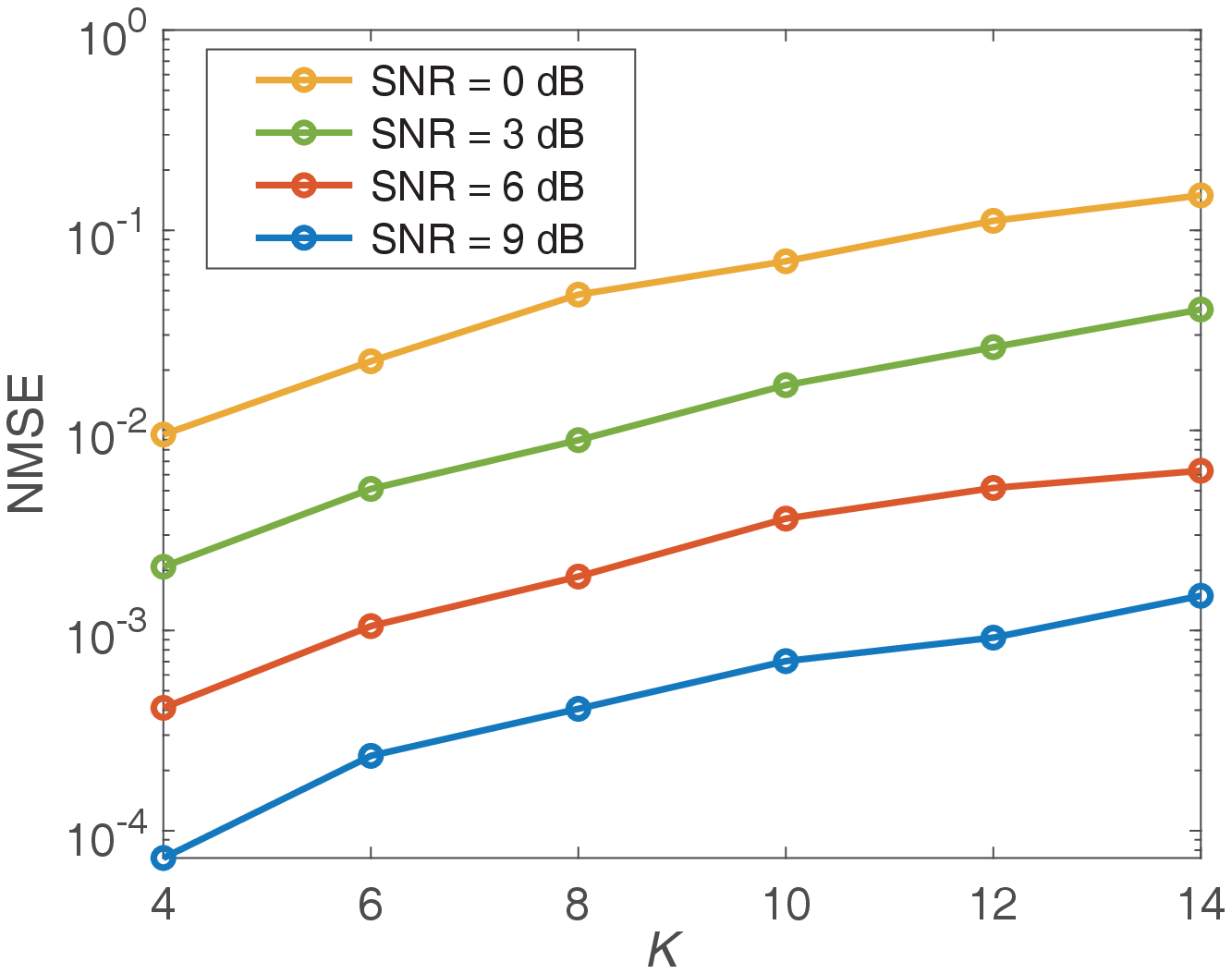}
\vspace{-0.5cm}
%\caption{The NMSE versus the number of sensor devices.}
\caption{The NMSE versus $K$.}
\vspace{-0.4cm}
\label{Fig:NMSEp}
\end{minipage}
\end{figure*}

\vspace{-0.5cm}
\section{Conclusion}

In this paper, we consider the joint optimization of transceiver and the RIS phase design for a RIS-aided AirComp system with imperfect CSI. The complicated optimization problem is equivalently transformed into a subproblem of robust design  and joint design of the system. We  solve the joint design problem under the worst-case robustness for CSI error model and obtain the closed-form solutions by utilizing efficient alternating algorithms and solving the Karush-Kuhn-Tucker conditions.
\vspace{-0.4cm}
\appendices
\section{Proof of Theorem 1}
By applying the Karush-Kuhn-Tucker (KKT) conditions \cite{KKT}, it is equivalent to solve the following conditions as
\begin{equation}\label{eq:p44}
\begin{aligned}
\min_{\Delta_{\mathbf{h}_{k}},\lambda_{k}}\quad &\mathcal{L}(\Delta_{\mathbf{h}_{k}},\lambda_{k})\\
&=-\left| m(\hat{\mathbf{h}}_{k}^{\rm H}+\Delta_{\mathbf{h}_{k}})\boldsymbol{v}_{k}t_{k}-1\right|^{2}+\lambda_{k}(\|\Delta_{\mathbf{h}_{k}}\|^2-\varepsilon^2)\\
{\rm s.t.}\quad &\frac{\partial\mathcal{L}(\Delta_{\mathbf{h}_{k}},\lambda_{k})}{\partial \Delta_{\mathbf{h}_{k}}^{*}}=0,\\
&\lambda_{k}(\|\Delta_{\mathbf{h}_{k}}\|^2-\varepsilon^2)=0,\\
&\|\Delta_{\mathbf{h}_{k}}\|^2-\varepsilon^2 \leq 0,\\
&\lambda_{k} \geq 0,\\
\end{aligned}
\end{equation}
where
%\begin{equation}\nonumber
%\mathcal{L}(\Delta_{\mathbf{h}_{k}},\lambda_{k})=-\left| m(\hat{\mathbf{h}}_{k}^{\rm H}+\Delta_{\mathbf{h}_{k}})\boldsymbol{v}_{k}t_{k}-1\right|^{2}+\lambda_{k}(\|\Delta_{\mathbf{h}_{k}}\|^2-\varepsilon^2),
%\end{equation}
%and
$\lambda_{k}$, for $k=1, 2, \cdots, K$, are the KKT multipliers.

Letting $\hat{t}_{k}=mt_k$, and from \eqref{eq:p44}, we have
\begin{align}\label{eq:deltahproof}
\frac{\partial\mathcal{L}(\Delta_{\mathbf{h}_{k}},\lambda_{k})\}}{\partial \Delta_{\mathbf{h}_{k}}^{*}}=&-\vert \hat{t}_{k} \vert^2\hat{\mathbf{h}}_{k}^{\rm H}\boldsymbol{v}_{k}\boldsymbol{v}_{k}^{\rm H}+\hat{t}_{k}^{*}\boldsymbol{v}_{k}^{\rm H}-\vert \hat{t}_{k} \vert^2\Delta_{\mathbf{h}_{k}}\boldsymbol{v}_{k}\boldsymbol{v}_{k}^{\rm H}\nonumber\\
&+\lambda_{k}\Delta_{\mathbf{h}_{k}}.
\end{align}
Forcing \eqref{eq:deltahproof} to zero, we obtain
\begin{align}
\Delta_{\mathbf{h}_{k}}=&(\vert \hat{t}_{k} \vert^2\hat{\mathbf{h}}_{k}^{\rm H}\boldsymbol{v}_{k}\boldsymbol{v}_{k}^{\rm H}-\hat{t}_{k}^{*}\boldsymbol{v}_{k}^{\rm H})(\lambda_{k} \mathbf{I}_{N}-\vert \hat{t}_{k} \vert^2\boldsymbol{v}_{k}\boldsymbol{v}_{k}^{\rm H})^{-1}\nonumber\\
=&-\hat{\mathbf{h}}_{k}^{\rm H}+(\lambda_{k}\hat{\mathbf{h}}_{k}^{\rm H}-\hat{t}_{k}^{*}\boldsymbol{v}_{k}^{\rm H})(\lambda_{k} \mathbf{I}_{N}-\vert \hat{t}_{k} \vert^2\boldsymbol{v}_{k}\boldsymbol{v}_{k}^{\rm H})^{-1}\nonumber\\
\overset{(a)}{=}&-\hat{\mathbf{h}}_{k}^{\rm H}+\lambda_{k}\hat{\mathbf{h}}_{k}^{\rm H}(\lambda^{-1}_{k}\mathbf{I}_{N}-\frac{-\lambda^{-2}_{k}\vert \hat{t}_{k} \vert^2\boldsymbol{v}_{k}\boldsymbol{v}_{k}^{\rm H}}{1-\lambda^{-1}_{k}\vert \hat{t}_{k} \vert^2\boldsymbol{v}_{k}^{\rm H}\boldsymbol{v}_{k}})\nonumber\\
&-\hat{t}_{k}^{*}\boldsymbol{v}_{k}^{\rm H}(\lambda_{k} \mathbf{I}_{N}-\vert \hat{t}_{k} \vert^2\boldsymbol{v}_{k}\boldsymbol{v}_{k}^{\rm H})^{-1}\nonumber\\
\overset{(b)}{=}&\frac{\lambda^{-1}_{k}\vert \hat{t}_{k} \vert^2\hat{\mathbf{h}}_{k}^{\rm H}\boldsymbol{v}_{k}\boldsymbol{v}_{k}^{\rm H}}{1-\lambda^{-1}_{k}\vert \hat{t}_{k} \vert^2\boldsymbol{v}_{k}^{\rm H}\boldsymbol{v}_{k}}-\frac{\lambda^{-1}_{k}\hat{t}_{k}^{*}\boldsymbol{v}_{k}^{\rm H}}{1-\lambda^{-1}_{k}\vert \hat{t}_{k} \vert^2\boldsymbol{v}_{k}^{\rm H}\boldsymbol{v}_{k}}\nonumber\\
=&\frac{\lambda^{-1}_{k}\vert \hat{t}_{k} \vert^2\hat{\mathbf{h}}_{k}^{\rm H}\boldsymbol{v}_{k}\boldsymbol{v}_{k}^{\rm H}-\lambda^{-1}_{k}\hat{t}_{k}^{*}\boldsymbol{v}_{k}^{\rm H}}{1-\lambda^{-1}_{k}\vert \hat{t}_{k} \vert^2\boldsymbol{v}_{k}^{\rm H}\boldsymbol{v}_{k}},\label{eq:deltahkp}
\end{align}
where $(a)$ uses the matrix inversion lemma as
\begin{equation}
(\mathbf{A}+\mathbf{u}\mathbf{v}^{\rm T})^{-1}=\mathbf{A}^{-1}-\frac{1}{1+\mathbf{v}^{\rm T}\mathbf{A}^{-1}\mathbf{u}}\mathbf{A}^{-1}\mathbf{u}\mathbf{v}^{\rm T}\mathbf{A}^{-1},
\end{equation}
and $(b)$ uses the fact that
\begin{equation}
\mathbf{x}^{\rm H}(\mathbf{A}+\tau\mathbf{x}\mathbf{x}^{\rm H})^{-1}=\frac{\mathbf{x}^{\rm H}\mathbf{A}^{-1}}{1+\tau\mathbf{x}^{\rm H}\mathbf{A}^{-1}\mathbf{x}}
\end{equation}
for any invertible matrix $\mathbf{A}$.

Substitute \eqref{eq:deltahkp} into the KKT complementary condition in \eqref{eq:p44}, i.e., $\lambda_{k}(\|\Delta_{\mathbf{h}_{k}}\|^2-\varepsilon^2)=0$ and solve this equality. By removing
unreasonable negative values, we get the value of $\lambda_{k}$.
% The detailed derivations go in (\ref{eq:lambdaproof}).

Similarly, we have the desired  worst-case MSE in \eqref{eq:msestar}.

\end{document}